\documentclass[aps,showpacs,prl,twocolumn]{revtex4}

\usepackage{amsmath,amssymb}
\usepackage{graphicx}
\usepackage{xcolor}

\newcommand{\ket}[1]{|#1\rangle}
\newcommand{\bra}[1]{\langle#1|}

\newcommand{\eq}{\begin{equation}}
\newcommand{\fine}{\end{equation}}

\begin{document}

\title{Realization and characterization of a 2-photon 4-qubit linear cluster state}
\author{Giuseppe Vallone$^{1,*}$, Enrico Pomarico$^{1,*}$, Paolo Mataloni$^{1,*}$,
Francesco De Martini$^{1,*}$, Vincenzo Berardi$^{2}$\\
$^{1}$Dipartimento di Fisica dell'Universit\'{a} ``La Sapienza'' and
Consorzio Nazionale Interuniversitario per le Scienze Fisiche della Materia,
Roma, 00185 Italy\\
$^{2}$Dipartimento Interateneo di Fisica, Universit\`{a} e Politecnico di
Bari and Consorzio Nazionale Interuniversitario per le Scienze Fisiche della
Materia, Bari, 70126 Italy}

\begin{abstract}
We report on the experimental realization of a 4-qubit
linear cluster state via two photons entangled both in polarization and linear momentum.
This state was investigated
by performing tomographic measurements and by evaluating an entanglement witness. 
By use of this state we carried out a novel nonlocality proof,
the so-called ``stronger two observer all versus nothing'' test of
quantum nonlocality.
\end{abstract}
\maketitle

Multipartite graph states and, in particular, cluster states, have been
recently introduced by Briegel and Raussendorf 
as a fundamental resource aimed at the linear optics one way
quantum computation \cite{brieg-rauss,brieg-rauss1}, 
and at the realization of important quantum
information tasks, such as quantum error correction and quantum
communication protocols \cite{schlingemann,cleve-hillery}.
Recently, the experimental feasibility of one way quantum
computation by four photon cluster states was demonstrated 
\cite{zeil-nature,quant-ph}. Besides the applications to quantum
computation, cluster states are powerful tools for perfoming nonlocality 
tests \cite{zeil-cluster,wein-cluster}. It is well known that the adoption
of an increasing number of internal qubits, i.e. in a higher dimensional
Hilbert space, leads to a stronger violation of local realism \cite{mermin-cabello}. 
Recently, a test demonstrating that nonlocality grows with the
number of internal degrees of freedom of the system, was indeed successfully carried
out by taking advantage of the peculiar properties of a $2$-photon hyperentangled
state \cite{gws}. 
It is worth noting that, at variance with the cluster states, 
hyperentangled, or double entangled states, are bi-separable and
do not represent genuine four-qubit entangled states.

	In this letter we report the experimental realization of a high
	fidelity $2$-photon $4$-qubit linear cluster state by a linear optical
	technique consisting of the entanglement of the
	polarization ($\pi $) and momentum ($\mathbf{k}$) degrees of freedom of one
	of the two photons belonging to an hyperentangled state. 
	The cluster state was analyzed 
	by quantum tomographic measurements and by 
	an entanglement witness method \cite{toth-gune,wein-cluster}. 
	By using this state, we performed a novel ``All-Versus-Nothing'' (AVN)
	test of nonlocality recently proposed by Cabello \cite{cluster-cabello}.
\begin{figure}[b]
\includegraphics[scale=.48]{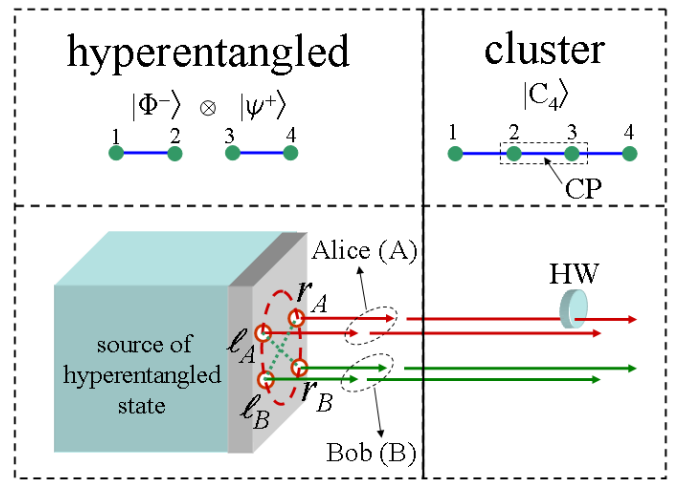}
\caption{Generation of the linear cluster state by a source of
polarization-momentum hyperentangled 2-photon state. The 
state $|\Xi \rangle =|\Phi^- \rangle \otimes |\protect{\psi^+} \rangle $
corresponds to two separate 2-qubit clusters. The $HW$ acts as a
Controlled-Phase (CP) thus generating the 4-qubit linear cluster $|C_{4}\rangle $. }
\label{source}
\end{figure}

As said, the starting point for the cluster state generation was the hyperentangled
state $|\Xi \rangle =\ket{\Phi^-}\otimes\ket{{\psi^+}}$, 
where
$|\Phi^- \rangle =\frac{1}{\sqrt{2}}
\left( |H\rangle _{A}|H\rangle_{B}-|V\rangle _{A}|V\rangle _{B}\right) $
and
$|{\psi^+} \rangle =\frac{1}{\sqrt{2}}(|r\rangle _{A}|\ell \rangle _{B}
+|\ell \rangle_{A}|r\rangle _{B})$. In the above equations
$H,V$ refer to the horizontal ($H$) and
vertical ($V$) polarizations and $\ell ,r$ refer to the left ($\ell $) or
right ($r$) paths of the photon $A$ (Alice) or $B$ (Bob) 
(see Fig. \ref{source}). 
The state $|\Xi \rangle $ is realized by a Spontaneous Parametric Down
Conversion (SPDC) method already described in details in other
papers \cite{PRA-hyper,PRA-pol}. A thin type I $\beta$-barium-borate
BBO crystal slab operating under the double 
(back and forth) excitation of a cw $Ar^{+}$
laser ($\lambda _{p}=364$ nm) generated the $\pi$-entangled state 
$|\Phi^- \rangle$, obtained by the superposition 
of two perpendicularly polarized SPDC cones emerging from the crystal at 
the degenerate wavelength $\lambda =728$ nm. 
The {\bf k}-entangled state $|{\psi^+} \rangle$ was realized by
selecting two pairs of correlated $\mathbf{k}$-modes, 
$r_A$-$\ell_B$ and $\ell_A$-$r_B$, belonging to the conical emission of the crystal. 
Because of the ``phase-preserving'' character of
the SPDC process, the relative phase between the two pair emissions was
set to the value $\phi =0$. By adoption of hyperentangled states
several AVN tests of quantum nonlocality were recently proposed \cite{AVN-teorici} and carried out \cite{hyper-AVN}.

In the present experiment the $2$-photon $4$-qubit linear cluster state
\eq
\begin{aligned}\label{cluster}
|C_{4}\rangle &=\frac{1}{2}\left( |Hr\rangle _{A}|H\ell \rangle
_{B}+|Vr\rangle _{A}|V\ell \rangle _{B}\right.  \\
&\qquad\qquad\left. +|H\ell \rangle _{A}|Hr\rangle _{B}-|V\ell \rangle _{A}|Vr\rangle
_{B}\right)   \\
&=\frac{1}{\sqrt 2}\left( \ket{\Phi^+}|r\rangle _{A}|\ell \rangle_{B}+
\ket{\Phi^-}|\ell\rangle _{A}|r\rangle _{B}\right)\,,
\end{aligned}
\fine
where $|\Phi ^{+}\rangle =
\frac{1}{\sqrt{2}}(|H\rangle _{A}|H\rangle_{B}+ |V\rangle _{A}|V\rangle _{B})$, 
was created by inserting in the $r_{A}$ (right-Alice) mode a zero 
order half wave plate ($HW$) with the optical
axis oriented along the vertical direction (see Fig. \ref{source}). The $HW$ left
the state $\ket{\Phi^-}\ket{\ell}_A\ket{r}_B$ unchanged, while the transformation
$\ket{\Phi^-}\ket{r}_A\ket{\ell}_B$ into $\ket{\Phi^+}\ket{r}_A\ket{\ell}_B$ also
transformed $\ket{\Xi}$ into $\ket{C_4}$.

Under the correspondence $|H\rangle _{A,B}\leftrightarrow |0\rangle _{2,1}$,
$|V\rangle _{A,B}\leftrightarrow |1\rangle _{2,1}$ and $|\ell \rangle
_{A,B}\leftrightarrow |0\rangle _{3,4}$, $|r\rangle _{A,B}\leftrightarrow
|1\rangle _{3,4}$, the state \eqref{cluster} is equivalent to the cluster
state 
\footnote{The state \eqref{cluster 4 photon} is equivalent to the 
linear cluster state
$|\phi_4\rangle=\frac{1}{2}
(|0000\rangle+|1100\rangle+|0011\rangle-|1111\rangle)$, 
generated in \cite{wein-cluster}, up to a $\sigma_{x}$ operation
on the third qubit.} 
expressed in the logical basis $|0\rangle $\ and\ $|1\rangle $
\begin{multline}
|\widetilde C_{4}\rangle =\frac{1}{2}\left( |0\rangle _{1}|0\rangle
_{2}|1\rangle _{3}|0\rangle _{4}+|1\rangle _{1}|1\rangle _{2}|1\rangle
_{3}|0\rangle _{4}\right. \\
\left. +|0\rangle _{1}|0\rangle _{2}|0\rangle _{3}|1\rangle _{4}-|1\rangle
_{1}|1\rangle _{2}|0\rangle _{3}|1\rangle _{4}\right)\,.
\label{cluster 4 photon}
\end{multline}
\begin{figure}[tbp]
\includegraphics[scale=.3]{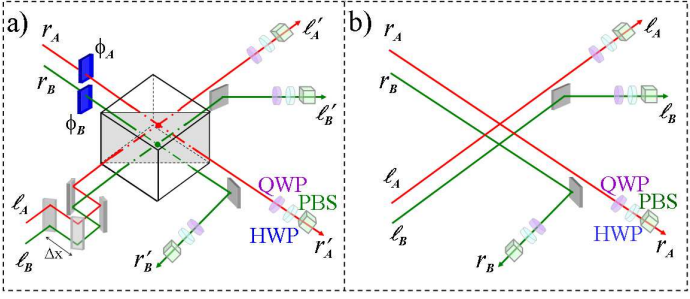}
\caption{Interferometer and measurement apparatus.
a) The mode pairs $r_A$-$\ell_B$ and $\ell_A$-$r_B$ are matched on the BS.
The phase shifters $\phi_A$ and $\phi_B$ (thin glass plates) are used for the
measurement of momentum observables.
The polarization analyzers on each of BS output modes are shown
(QWP/HWP=Quarter/Half-Wave Plate,  PBS=Polarized Beam Splitter).
b) Same configuration as in a) with BS and glasses removed.}
\label{BS}
\end{figure}

It is worth stressing that the insertion of $HW$ represents a ``local''
operation in the sense that it acts on photon $A$ only, while it is
``nonlocal'' for the two qubits associated to photon $A$ itself. 
Indeed, the $HW$ operates
as a Controlled Phase (CP) between the target qubit $2$ and the control qubit $3$
(i.e. the polarization and the momentum degree of freedom of photon $A$),
thus entangling the four qubits together.
Moreover this operation does not require any kind of post-selection.

Let's consider the measurement setup shown in Fig. \ref{BS}a).
The mode pairs 
$r_A$-$\ell_B$ and $\ell_A$-$r_B$
are there spatially and temporally superimposed by means of a $50\%$ beam splitter
(BS). The optical path delay $\Delta x$ can be simultaneously 
changed for both $\ell_A$ and $\ell_B$ modes by using a trombone mirror assembly.
The null value delay ($\Delta x=0$) corresponds to the
exact superposition on the BS between 
$r_A$-$\ell_B$ and $\ell_A$-$r_B$, 
i.e. when the right ($r$) and left ($\ell$) optical paths are equal \cite{PRA-hyper}.
The two thin glass plates inserted on the right modes ($\phi_A$ and $\phi_B$ in Fig. \ref{BS}a)) and the BS 
transform the input states in the following way:
$\frac{1}{\sqrt2}(\ket{\ell}_i+e^{-\text{i}\phi_i}\ket{r}_i)\rightarrow\ket{\ell'}_i$,
$\frac{1}{\sqrt2}(\ket{\ell}_i-e^{-\text{i}\phi_i}\ket{r}_i)\rightarrow\ket{r'}_i$ with $i=A,B$.
Note that these are single photon transformations: in fact the single BS showed in Fig. \ref{BS}a) is equivalent 
to two BS's, one for each (A or B) particle. The reason why we used the single BS
apparatus resides on its higher phase stability.

The photons associated with the BS output modes $\ell'_A$, $r'_B$, $\ell'_B$ and $r'_A$
are analyzed each by a quarter-wave plate (QWP), half-wave plate (HWP), 
a polarizing beam splitter (PBS) and detected by single photon avalanche detectors. 
Two thin glass plates on modes $r_A$ and $r_B$ are properly set 
for measuring momentum observables.
The analysis setup shown in Fig. \ref{BS}b) is obtained 
from Fig. \ref{BS}a) by removing the interferometric apparatus and 
allows the measurement
of several relevany observables that will be introduced later in the paper. 
\begin{figure}[tbp]
\includegraphics[scale=.4]{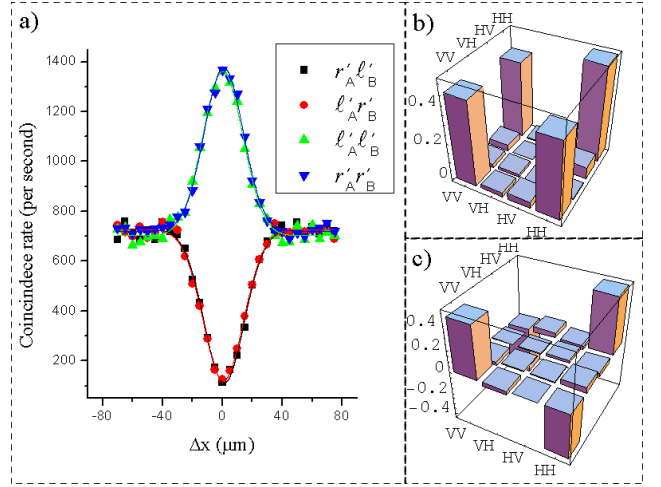}
\caption{State characterization. a) Coincidence rates versus path delay $\Delta x$
showing the interference pattern between the two pairs $r_A$-$\ell_B$ and $\ell_A$-$r_B$.
The dip(peak) FWHM and the coherence time ($\sim150fsec$) of the photons are determined by
the bandwidth ($6$nm) of the interference filter used.
b) and c) Tomographic reconstructions (real parts) of the polarization states corresponding
to $|\Phi ^{+}\rangle$ and $|\Phi ^{-}\rangle$ respectively. The imaginary components are negligible.
Typical uncertainties are $0.006$ for the higher terms ($\ket{HH}\bra{HH}$, $\ket{HH}\bra{VV}$,
$\ket{VV}\bra{HH}$ and $\ket{VV}\bra{VV}$) and $0.003$ for the other terms.} 
\label{tomo}
\end{figure}

We characterized the state \eqref{cluster} by
measuring the interference between the mode pairs
$r_A$-$\ell_B$ and $\ell_A$-$r_B$ as a function of the delay $\Delta x$.
The dip-peak graph ($88\%$ average visibility) for $H$ polarized photons,
corresponding to the ${\bf k}$-entangled state
$\ket{\psi^+}$,
is shown in Fig. \ref{tomo}a). Similar results are obtained for
$V$ polarized photons with dips and peaks flipped 
\footnote{The cluster state \eqref{cluster} can be also written as
$\ket\Xi=\frac1{\sqrt2}(\ket{H}_A\ket{H}_B\ket{\psi^+}+\ket{V}_A\ket{V}_B\ket{\psi^-})$
where
$\ket{\psi^-}=\frac1{\sqrt2}(\ket{r}_A\ket{\ell}_B-\ket{\ell}_A\ket{r}_B)$,
then dips and peaks are flipped for $V$ photons.}.
By removing the BS (Fig. \ref{BS}b)),
we performed a quantum tomographic analysis on the mode sets
$r_{A}$-$\ell _{B}$ and $\ell _{A}$-$r_{B}$, 
corresponding to the $\pi$-entangled states
$\ket{\Phi^+}$ (Fig. \ref{tomo}b)) and $\ket{\Phi^-}$ (Fig. \ref{tomo}c)) respectively.
The tomographic reconstructions were obtained by the ``Maximum Likelihood Estimation'' method described in \cite{01-jam-mea}.
The corresponding fidelities are $F_{\ket{\Phi^+}}=0.9068\pm0.0035$ 
and $F_{\ket{\Phi^-}}=0.9131\pm0.0032$. 
Note that the path interference measurement shown
in Fig. \ref{tomo}a) demonstrates the quantum superposition between the two
states
$\ket{\Phi^+}|r\rangle _{A}|\ell \rangle_{B}$ and 
$\ket{\Phi^-}|\ell\rangle _{A}|r\rangle _{B}$
of Fig. \ref{tomo}b) and \ref{tomo}c), leading to the linear cluster state
\eqref{cluster}.

\begin{figure}[tbp]
\includegraphics[scale=.29]{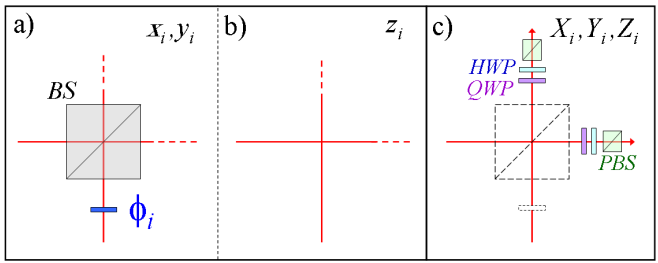}
\caption{Measurement setup for
momentum \big(a),b)\big) and polarization \big(c)\big) observables 
for photon $i$ ($i$=$A,B$). 
By the a) setup we measure
$x_i$ ($\phi_i=0$) and $y_i$ ($\phi_i=\frac\pi2$),
while the b) setup is used for measuring $z_i$. By the c) setup we measure 
$X_i$ ($\theta_Q=\frac\pi4$; $\theta_H=\frac18\pi,\frac38\pi$), 
$Y_i$ ($\theta_Q=0$; $\theta_H=\frac18\pi,\frac38\pi$) and
$Z_i$ ($\theta_Q=0$; $\theta_H=0,\frac\pi4$),
where $\theta_{H(Q)}$
is the angle between the $HWP$($QWP$) optical axis and the vertical direction. 
The polarization analysis is performed contextually to $x_i$, $y_i$ (i.e. with BS and glass)
or $z_i$ (without BS and glass), as shown by the dotted lines for BS and glass in c).
}
\label{operators}
\end{figure}

The genuine multipartite 4-qubit entanglement was verified 
by measuring the entanglement witness \cite{toth-gune}
\begin{multline}
\mathcal W=\frac12\left[
4\openone-Z_AZ_B-Z_Ax_Ax_B+X_Az_AX_B\right.\\
\left.+z_Az_B-x_AZ_Bx_B-X_AX_Bz_B\right]
\label{ent-wit}
\end{multline}
where upper cases refer to polarization operators
\eq\label{pol}
\begin{aligned}
&Z_i=\ket{H}_i\bra{H}-\ket{V}_i\bra{V}\\
&Y_i=i\ket{V}_i\bra{H}-i\ket{H}_i\bra{V}\\
&X_i=\ket{H}_i\bra{V}+\ket{V}_i\bra{H}
\end{aligned}
\qquad \qquad i=A,B
\fine
and lower cases refer to momentum operators
\eq\label{mom}
\begin{aligned}
&z_i=\ket{\ell}_i\bra{\ell}-\ket{r}_i\bra{r}\\
&y_i=i\ket{r}_i\bra{\ell}-i\ket{\ell}_i\bra{r}\\
&x_i=\ket{\ell}_i\bra{r}+\ket{r}_i\bra{\ell}
\end{aligned}
\qquad \qquad i=A,B
\fine
The experimental setups for measuring the polarization \eqref{pol} and momentum
\eqref{mom} observables  for either Alice or Bob photon
are shown in Fig.  \ref{operators}.
Note that the eigenvectors of $x_i$ and $y_i$ can be written in the form $\frac{1}{\sqrt2}(\ket{\ell}_i\pm e^{-\text{i}\phi_i}\ket{r}_i)$.
Those states can be discriminated, as previously explained, by the glass plates and the BS.

The expectation value of $\mathcal W$ is positive for any
separable state 
(for instance it is equal to 1 for the hyperentangled state $\ket\Xi$),
whereas its negative value detects
4-party entanglement close to the cluster state \eqref{cluster}.
A perfect cluster state gives $-1$ as expectation value.

The experimental values of the observables of eq. \eqref{ent-wit}
are shown in Table \ref{table}. The non perfect correlations
were due to the impurity of the states $\ket{\Phi^+}$ and $\ket{\Phi^-}$,
as well as to imperfections in the polarization and momentum analysis 
devices.
The resulting experimental value of $\mathcal W$ is 
\eq
\text{Tr}[\mathcal{W}\rho_{exp}]=-0.6843\pm0.0094\,,
\fine
thus demonstrating the genuine multipartite entanglement of 
our cluster state, whose $\rho_{exp}$ represents the experimental density matrix.

From the projector-based entanglement witness \cite{toth-gune}
$\widetilde {\mathcal W}=\frac12-\ket{C_4}\bra{C_4}\,,$
we could obtain information about the fidelity $F_{\ket{C_4}}$ of the state  
through the equation $F_{\ket{C_4}}=\frac12-\text{Tr}[\widetilde W\rho_{exp}]$.
As shown in \cite{toth-gune}, 
the following relation holds between 
$\mathcal W$ and $\widetilde{\mathcal W}$: 
$\mathcal W-2\widetilde{\mathcal W}\geq0$.
Hence the lower bound of the experimental fidelity $F_{\ket{C_4}}$ is:
\eq
F_{\ket{C_4}}\geq\frac12-\frac12\text{Tr}[\mathcal{W}\rho_{exp}]\geq0.84\,,
\fine
giving a further evidence of the cluster generation.
\begin{table}[t]
\begin{tabular}{ccccc}
\hline\hline
\hspace{.6cm}Observable\hspace{.6cm}
&\hspace{1cm}Value\hspace{1cm}
&\hspace{.2cm}$\mathcal W$\hspace{.2cm}
&\hspace{.2cm}$S$\hspace{.2cm}
&\hspace{.2cm}$C$\hspace{.2cm}\\
\hline
$Z_AZ_B$       & $+0.9283\pm0.0032$ &$\checkmark$ &             &\\
$Z_Ax_Ax_B$    & $+0.8194\pm0.0049$ &$\checkmark$ &             &\\
$X_Az_AX_B$    & $-0.9074\pm0.0037$ &$\checkmark$ &             &$\checkmark$\\
$z_Az_B$			 & $-0.9951\pm0.0009$ &$\checkmark$ &             &$\checkmark$\\
$x_AZ_Bx_B$    & $+0.8110\pm0.0050$ &$\checkmark$ &             &$\checkmark$\\
$Z_Ay_Ay_B$    & $+0.8071\pm0.0050$ &					    &             &$\checkmark$\\
$Y_Az_AY_B$    & $+0.8948\pm0.0040$ &             &             &$\checkmark$\\
$X_AX_Bz_B$    & $+0.9074\pm0.0037$ &$\checkmark$ &$\checkmark$ &$\checkmark$\\
$Y_AY_Bz_B$    & $-0.8936\pm0.0041$ &             &$\checkmark$ &$\checkmark$\\
$X_Ax_AY_By_B$ & $+0.8177\pm0.0055$ &             &$\checkmark$ &\\
$Y_Ax_AX_By_B$ & $+0.7959\pm0.0055$ &					    &$\checkmark$ &\\
\hline
\hline
\end{tabular}
\caption{Experimental values of the observables used for
measuring the entanglement witness $\mathcal W$ and the 
expectation value of $S$ on the cluster state $|C_{4}\rangle$. 
The third column ($C$) refers to the control measurements needed to verify that
$X_A$, $Y_A$, $x_A$, $X_B$, $Y_B$, $y_B$ and $z_B$
can be considered as elements of reality.
Each experimental value corresponds to a measure lasting an average time of 10 sec.
In the experimental errors we considered the poissonian statistic and the uncertainties 
due to the manual setting of the polarization analysis wave plates.}
\label{table}
\end{table}

Finally, we tested the nonlocal character of our cluster state by
using the ``stronger two observer AVN'' proof
of local realism, recently introduced in \cite{cluster-cabello}.
It is based on the following eigenvalue equations:
\begin{subequations}
\begin{align}
X_Az_AX_B\ket{C_4}&=-\ket{C_4}\\
z_Az_B\ket{C_4}&=-\ket{C_4}\\
x_AZ_Bx_B\ket{C_4}&=+\ket{C_4}\\
Z_Ay_Ay_B\ket{C_4}&=+\ket{C_4}\\
Y_Az_AY_B\ket{C_4}&=+\ket{C_4}\\
X_AX_Bz_B\ket{C_4}&=+\ket{C_4}\\
Y_AY_Bz_B\ket{C_4}&=-\ket{C_4}\\
X_Ax_AY_By_B\ket{C_4}&=+\ket{C_4}\\
Y_Ax_AX_By_B\ket{C_4}&=+\ket{C_4}
\end{align}
\end{subequations}
The first seven equalities demonstrate that the local observables
$X_A$, $Y_A$, $x_A$, $X_B$, $Y_B$, $y_B$ and $z_B$ are elements
of reality in the EPR sense \cite{EPR}.
The last four equalities are used in the AVN proof through
the following quantum mechanical expectation value of the cluster state
\eqref{cluster}:
\begin{multline}\label{AVN}
\langle S\rangle=\bra{C_4}X_AX_Bz_B-Y_AY_Bz_B\\
+X_Ax_AY_Ay_B+Y_Ax_AX_Ay_B\ket{C_4}=4
\end{multline}
In any local realistic theory based on the
previously defined elements of reality, the upper bound
of the expected value for eq. \eqref{AVN} is $2$.

From the experimental values given in Table \ref{table}
we obtain 
\eq
\text{Tr}[S\rho_{exp}]=3.4145\pm0.0095\,,
\fine
which violates the classical bound by 148 standard deviations.
Note that this result provides another enhanced discrepancies between the quantum
versus classical predictions (4 versus 2) with respect to the standard CHSH inequality ($2\sqrt2$ versus $2$)
\cite{gws}.

In summary, in this letter we have presented the experimental realization of a high
fidelity linear cluster state consisting of four entangled qubits
by adoption of $2$-photon polarization-momentum hyperentanglement
within a linear optical method.
The cluster state was generated by applying a CP gate between the polarization
and momentum qubits of one photon of the hyperentangled state.
The genuine entangled character of the cluster state
was experimentally demonstrated. Its nonlocal behaviour was also tested by a novel AVN 
quantum mechanical test
proposed for $2$-photon linear cluster state.

Other kinds of cluster states can be easily produced
by the same technique presented here. Apart for the relevance of these
states for fundamental physics, two photon cluster states may be good
candidates to realize important quantum information tasks
because of their high purity and the relatively high generation rate. 
Whether or not these states may also represent a useful resource 
for linear optics quantum computation is as yet unclear.
In fact our method could be used in probabilistic quantum computation
with the advantage of high counting rates. Moreover, two $\ket{C_4}$ states
generated by the same laser source could be linked together by a suitable CP
gate to form an 8-qubit linear cluster state.

Thanks are due to Ad\'an Cabello for useful discussions and Marco Barbieri for
his contribution in planning the experiment. This work was supported by the
FIRB 2001 (Realization of Quantum Teleportation and Quantum Cloning by the
Optical Parametric Squeezing Process) and PRIN 2005 (New perspectives in
entanglement and hyper-entanglement generation and manipulation) of MIUR
(Italy).
\\
\\
*Electronic address: http://quantumoptics.phys.uniroma1.it/

\end{document}